\documentclass[5p,times]{elsarticle}

\usepackage{subfigure}
\usepackage{amsmath}
\usepackage{amssymb}
\usepackage{tabularx}
\usepackage[table]{xcolor}
\usepackage{adjustbox}
\usepackage{array}
\usepackage{pgfplots}
\usepackage{bchart}
\usepackage{times}
\usepackage{multirow}
\usepackage{cancel}
\usepackage{tikz}
\usepackage{url}
\usepackage{hyperref}
\usepackage{graphicx}
\usetikzlibrary{positioning,calc,decorations.pathreplacing,arrows.meta,bending}

\usepackage{mathtools}
\DeclarePairedDelimiter\ceil{\lceil}{\rceil}
\DeclarePairedDelimiter\floor{\lfloor}{\rfloor}

\usepackage{algorithm}
\usepackage{algpseudocode}
\usepackage{multicol}

\journal{arxiv.org}

\begin{document}
	
	\begin{frontmatter}
		
		
		
		\title{Precomputed Dominant Resource Fairness}
		
		
		\author[1]{Serdar Metin\corref{cor1}}
		\ead{balikakli@gmail.com}
		
		\cortext[cor1]{Corresponding author}
		
		\address{Alçakdam Yokuşu Sokak 9/7 Cihangir Beyoğlu İstanbul}

		\begin{abstract}
		Dominant Resource Fairness (DRF) is a resource allocation scheme, addressing the problem of fair allocation of multiple resource types, among users with heterogeneous demands for different types of resources. Based on Max-min Fairness, which is a well established algorithm in the literature for allocating resources in the single resource type scenarios, DRF generalises the main principle of maximising the minimum share given to any user, to the multiple resource case. It has a number of desirable properties, such as Sharing Incentive, Envy-Freeness, Pareto Efficiency, and Strategy Proofness, that makes it preferable over alternatives, and as such, it is widely adopted in distributed systems. In the present study, we revisit the original study, and analyse the structure of the algorithm in closer view. The main contribution is an alternative algorithm, which approximates the DRF allocation in linear time, significantly improving over the original algorithm's pseudo-polynomial runtime. We name the new algorithm Precomputed Dominant Resource Fairness (PDRF), after its main working principle. We provide bounds for approximating the original algorithm, and critically evaluate aforementioned properties comparing DRF and PDRF over them. It turns out that there is an adversarial strategy, which forces DRF to unfairly allocate more resources to users who submit inflated demands. PDRF, on the other hand, favouring smaller demands, penalises such adversarial behaviour and creates incentive for users to submit their minimal demands.
		\end{abstract}
		
		
		\hyphenation{mat-ched}
		
	\end{frontmatter}
	

	\section{Introduction}
	\label{introduction}
	
	Dominant Resource Fairness (DRF) is a resource allocation scheme, which aims to provide fairness of allocation among users with heterogeneous demands, in a time-sharing system with multiple resource types, such as computing clouds \citep{ghodsi2011dominant}. The services offered by these systems may include processing power, memory, and bandwidth, all of which are expected to be distributed fairly among the clients, according to some definition of fairness.
	
	Although the problem has been well-studied for single resource types \citep{keshav1997engineering,bertsekas1992data}, prior to DRF, less attention was paid to the multi-resource scenarios. The general approach was to adopt well established solutions for single resource allocation, such as Max-min Fairness (MF) and its derivatives, for a fixed bundle with differing amounts of resources from each resource type; e.g. standardised virtual machines. But this approach leads to inefficiencies, since the needs of different users are heterogeneous, where one needs predominantly one resource type, the other needs another. In the end, users ended up with unused resources that come in the bundle with the resources they need, where at the same time there were other users in the system in demand of those unused resources.
	
	Built on the MF premise, which aims to maximise the minimum share given to any user in a single resource setting, DRF aims to generalise the scheme by maximising the minimum relative share given to any user, over her dominant share; which in turn is defined for each user as the highest ratio of her demanded resources to their total reserve in the system (e.g. the ratio of demanded memory to the total available memory). The rest of the resources is allocated, for each user, in their fixed proportions to the user's dominant share, which is known as the \textit{Leontief Preferences} in the economy literature.
	
	DRF has acquired wide adoption in the distributed systems and cluster computation because of its properties, such as being \textit{Pareto Efficient}, \textit{Envy Free}, \textit{Strategy Proof}, and creating \textit{Sharing Incentive}.
	
	In the present study, we analyse the DRF algorithm as introduced in \cite{ghodsi2011dominant} in closer view, and based on the observations gathered thereby, introduce a new algorithm to approximate the DRF allocation in fewer steps. We call this new algorithm Precomputed Dominant Resource Fairness (PDRF), after its main working principle of precomputing the main allocation loop of DRF, and assigning resources to users without going through the tedious iterations of the loop.
	
	One study that also needs to be mentioned at the onset, is \cite{parkes2015beyond}, the method and the findings of which closely resemble to that of the present study, with some key differences that allow us to implement our newly proposed algorithm. The analysis method and the algorithm they propose, the Extended Dominant Resource Fairness (EDRF) are similarly based on precomputing the final allocation of DRF, but they modeled the underlying linear programming problem, rather than the algorithm itself. They construct a theoretical framework that allows them to provide more rigorous proofs of DRF properties, which is the main objective of the article. This framework is also used in the present study to establish the bounds for approximation of PDRF to DRF.
	
	The rest of this paper is organized as follows: In the next section, we briefly review the literature for related work. In Section \ref{models}, we review the foundational DRF and EDRF models. We introduce our core analytical method in Section \ref{cycles}, followed by the presentation of the PDRF algorithm in Section \ref{pdrf}. Section \ref{mathematics} provides the formal mathematical model and establishes the tight approximation bounds of PDRF relative to DRF. In Section \ref{experiments}, we present our empirical evaluation results, and departing from the observations gathered there, Section \ref{heuristic} introduces a clarifying heuristic method to allocate residual resources. Section \ref{performance} provides a comprehensive computational complexity analysis of the algorithmic speed-up factor. Section \ref{discussion} discusses game-theoretic strategy-proofness limitations (\ref{strategy}) alongside multi-resource weighting policies (\ref{weighting}). Finally, we conclude and outline future research directions in Section \ref{conclusion}.

	\section{Related Work}
	\label{related}
	
	DRF is first proposed in \cite{ghodsi2011dominant}, where they introduce the new notion of establishing an MF allocation among \textit{dominant shares}, rather than a fixed resource bundle of different resource types. They also provide an algorithm to carry out the distribution, and produce empirical data to demonstrate the performance of the algorithm. In their theoretical sections, they compare the virtues of the newly introduced allocation scheme to the ones existing in the economy literature, for this is as much a problem in the field of economy as it is in the field of computer science. Their results clearly demonstrate that DRF possesses various advantages over its alternatives.
	
	The model is challenged in \cite{dolev2012no}. They developed an alternative fairness notion, which they refer to as \textit{no justified complaints}, based on the assumption that the fairness of distribution of resources should be assessed from a global system point of view, considering the bottleneck property of resources. A resource is a bottleneck resource, if the availability of this resource is limited and there is a race condition for accessing it. They developed the \textit{Bottleneck Based Fairness} (BBF) algorithm to implement this paradigm as an alternative to DRF, with relative advantages of each algorithm in different use cases.
	
	Not only challenged, but DRF has also been adopted, generalised, contextualised, or extended to different cases. For example, Gutman and Nisan \cite{gutman2012fair}, introduced an economical framework to compare and contrast the two systems described above. While doing so they also critically assessed both approaches from an economic point of view. Joe-Wong and colleagues \cite{joe2013multiresource} investigates the trade-off between fairness and efficiency, providing a unifying framework with two families of fairness functions, DRF being a special case in one of them. Kash \cite{kash2014no} takes a scenario in which the complete information on demands is not available, since users arrive and leave the system, and develops a Dynamic DRF (DDRF) to address the problem. The DDRF algorithm is improved for its performance in \cite{li2015note}.
	
	Finally, DRF has been reformulated by Parkes and colleagues \cite{parkes2015beyond} to EDRF, in order to study its properties more rigorously. EDRF is also a solution to an abstract case, where the resources, as well as user demands are well-divisible, i.e. fractions of demands can be allocated to users. Unfortunately, this is not widely applicable to real-life scenarios, where a fraction of a task does not provide any added utility to the users. In the same article, they also introduce Sequential Min-max Algorithm to account for the indivisiblity of resources, and provide an alternative solution to the original question. One more contribution of EDRF is that it addresses the situations where there are users who submit zero demands for one or more resources.
	
	We now proceed with presenting DRF and EDRF models, as well as and in relation to our own, in the following section.
	
	\section{Related Models}
	\label{models}
	
	In this section we review DRF and EDRF in further detail for their close relevance to our work. The notions introduced in this section will be used to construct our own model in the following sections.
	
	\subsection{Dominant Resource Fairness}
	\label{drf}
	
	The DRF allocation scheme is build on the concept of dominant shares. The dominant share for each user is defined as the maximum of the fraction of her demands for different resource types to their respective reserves in the system. As exemplified in \cite{ghodsi2011dominant}, in a system with $9$ CPUs and $18$ GB memory, the dominant share, $ds$ for user A, with the demand vector $\left\langle 1, 4 \right\rangle $ is memory, since the \textit{fractional demands}, $fd$, of the user are $1/9$ and $4/18 = 2/9$, respectively. Similarly, the $ds$ of a user with the demand vector $\left\langle 3, 1 \right\rangle$ is for CPU, since $fd$'s are $3/9$ and $1/18$, respectively. The question then is, what is a fair allocation between the users, whose demands for the resources are defined as such.
	
	If the number of tasks allocated to users A and B is denoted by $x$ and $y$, respectively, the DRF allocation is defined as the solution to the optimising problem below \citep[p. 4]{ghodsi2011dominant}:
	
	\begin{align*}
		&& &&\max(x,y)		&&  		&&				&&\text{Maximise allocation} && &&\\
		&& &&x + 3y			&& \leq	 &&9				&&\text{CPU Constraint} && &&\\
		&& &&4x + y			&& \leq	 &&18			&&\text{Memory Constraint} && &&\\
		&& &&\frac{2x}{9}	&&   =	 &&\frac{y}{3}	&&\text{Equalise Dominant Shares} && &&
	\end{align*}\footnote{The formulation above is the verbatim reproduction from \cite{ghodsi2011dominant}. The expression $\max(x,y)$, together with the equality $\frac{2x}{9}=\frac{y}{3}$, is intended to represent the DRF objective of maximising the minimum absolute allocated dominant share. Equivalently, this objective can be written explicitly as
	$\max_{x,y}\min\left\{\frac{2x}{9},\frac{y}{3}\right\}$, subject to the stated resource constraints.}
	
	Designed to solve this problem, DRF applies MF principle to maximise the minimum share, on the users' $ds$'s by looping on the total allocated $ds$'s of users, taking the least allocated at each iteration, and allocating one unit of her demand vector, assumed to correspond to one \textit{task}, until one of the resources is exhausted. 
	
	Although not explicitly stated, in addition to this mechanism, DRF utilises a tie-breaking policy to take the user with the higher $ds$, in instances when more than one user is allocated the same minimum share. This we see at the example given in the article, in the allocation process for the above mentioned case.
	
	In the initial state, where both user's allocated $ds$ is $0$, the algorithm starts assigning one task to user B, which updates the state to $0$ allocated $ds$ for A, and $1/3$ for B. Since now the least allocated $ds$ belongs to A, the algorithm picks her and assigns one task, updating the state to $2/9$ w.r.t. $1/3 = 3/9$. Still A has the lower allocated $ds$, so one more task is allocated to her, updating to $4/9$ w.r.t. $3/9$, and pushing B to the position of the user with the least allocated $ds$. Thusly the algorithm proceeds with B, and continuing further in the similar fashion, assigns $2/3$ dominant shares to each user, at which point the CPU is saturated and the algorithm stops.
	
	As seen in the example above, DRF adopts progressive filling strategy, by aiming to keep the growth of allocation among different agents at the same pace. DRF reduces to MF, and also to BBF \citep{dolev2012no}, in cases where all users' dominant resources are the same.
	
	\subsection{Extended Dominant Resource Fairness}
	\label{edrf}
	
	The study by \cite{parkes2015beyond} introduces an alternative view of DRF. Instead of an iterative algorithm, they aim to precompute the resulting distribution, with a linear program. For doing so, they employ two notions in the setting. First, in a given user demand vector, the ratio between the entries for different resources is fixed. Second, the operating principles of progressive filling algorithm dictates that the pace at which the allocation for dominant shares of different users grow should be equated.
	
	Departing from these notions, it is possible to normalise each demand vector to its dominant share, and abstract that they will all be allocated simultaneously. In other words, if we scale all the dominant shares to $1$, scale all other fractional shares accordingly, we end up in a setting in which at each step, every user takes $1$ normalised allocation, and fairness among the dominant shares is kept intact.
	
	The question then is, what is the maximum number of such allocations that can be made before depleting one of the resources? The answer starts with normalising the vector of fractional demands,$fd_{ij}$ to the dominant share, $ds_i$, which is denoted $d_{ij}$ and defined more explicitly as:
	
	\[
	d_{ij} = \frac{fd_{ij}}{ds_i}
	\]
	
	As such, each user's dominant share is equated to $1$ and the question is reduced to find the maximum scaling factor $x$ such that:
	
	\[
	\sum_{i = 1}^{n} x \cdot d_{ij} \leq 1, \forall r \in R
	\]
	
	\noindent where $R$ is the set of resources. The answer is obtained simply by extracting $x$ out, and passing the summation to the other side of the inequality; obtaining:
	
	\[
	x = \frac{1}{\max_r \sum_i d_{ij}}
	\]
	
	At this point, the authors allow themselves a rather unrealistic abstraction of both well-divisible resources and well-divisible tasks. But although unrealistic, it is not unreasonable, since the abstraction lends a great ease for the mathematical analysis of DRF properties, without loss of generality.
	
	Before explicitly describing the exact allocation of each resource to each user, they go on to introduce a weighting mechanism, and then suffice to describe the assignment process in the pseudocode for the weighted version, of which the unweighted version is a special case where all the users are weighted equally for all the resources. Yet, it is rather straightforward that the allocation of resource $r$ to user $i$ is given by $x \cdot d_{ij} \cdot r$.
	
	The rest of the article is rather notationwise involved and out of the scope of the present study, since we are not investigating the DRF properties, but taking them as they are in the literature, and aiming to improve on the algorithm. They also present another algorithm for dropping the well-divisibility assumption for tasks, but it is an iterative algorithm and precomputing is not involved, leaving us no bases for comparison. Thus, in the present study, we will suffice to present the model to this extent.
	
	One important contribution of the article is to point out the fact that the resulting distribution cannot guarantee Pareto Efficiency, as such. In order for an allocation to be Pareto Efficient, it should not be possible to increase the utility of one user without decreasing another. However, the algorithm stops when a resource is depleted; yet, there may be tasks who do not need the depleted resource, and the remaining resources may be sufficient to further increase those tasks' utilisation, without decreasing any other task's. To address this issue, the algorithm takes multiple allocation \textit{rounds}, each one ending with the depletion of one resource, and the next one starting \textit{after} removing the saturated demands, until all \textit{demands} are saturated.

	\section{Cyclic Structure of the DRF Main Loop}
	\label{cycles}
	
	We will shortly move on to introducing PDRF, but before doing so, we will present a key observation that led to the development of the algorithm. It also introduces, what we believe is a simpler and more intuitive approach to precomputing the resulting DRF allocation for given sets of demands and resources.
	
	As indicated, the progressive filling algorithm, which DRF is based upon, operates on the principle of keeping the rate of growth among the tasks of each user at the same pace. As a consequence, the number of iterations at which a given task is allocated one unit of its demand vector is \textit{inversely} proportional to its dominant share, the tasks with lower dominant shares being visited more often, where the task with the highest dominant share, $ds^*$, is visited the least.
	
	Departing from this observation, it is possible to identify \textit{cycles} within the main allocation loop. The structure of the cycle is dependent on the distribution of the demands and the preferred tie-breaking policy in case of equal dominant shares. For example, the original DRF algorithm takes the demand with $ds^*$ first, in cases where the users have been allocated the same amount of resources thereto, as seen in the example in Section \ref{drf}. The user with $ds^*$, User B, with $ds = 1/3$, takes the first iteration, followed by two iterations of the User A with $ds = 2/9$. One more allocation to User B followed by, this time, a single allocation to User A completes the cycle and exhausts the reserve of the available CPU; at which point the algorithm returns.
	
	Note that at the completion of a cycle, perfect fairness is achieved, all users getting the same amount of dominant shares. On the other hand, it is fair to presume that in real-life scenarios this rarely coincides with the exhaustion of resources.
	
	In a system of $n$ users, the length of the cycle and the number of occurrences for each individual demand are given by 

	\[ \lvert l \rvert = \sum_{i = 1}^{n} \frac{ds^*}{ds_i}\] 

	\noindent where $\lvert l \rvert$ denotes the length of the cycle, and we assume all $ds^*/ds_i$ values are integers. The number of occurrences of each demand within the cycle is given by the $i$-$th$ term of the summation, for user $i$.
	
	The ratios $ds^* / ds_i$ are still useful, in cases where integer-ratio assumption does not hold. In those cases, further \textit{subcycles} emerge, characterised by the single occurrence of $ds^*$, followed by all other demands occurring in the number of the ratio $ds^*$ to $ds_i$ (i.e. $ds^*/ds_i$), rounded up or down. The most frequently occurring allocation count is the nearest integer to the ratio.
	
	\[
	\lvert l' \rvert = \sum_{i=1}^{n} \floor*{\frac{ds^*}{ds_i}}
	\]
	
	\noindent where $\lvert l' \rvert$ denotes the length of the shortest \textit{possible} subcycle. The other subcycles are characterised by the occurrence of one or more of the lower dominant share demands equal to the ratio rounded up, the frequency of which is determined by the decimal part of the ratio $ds^*/ ds_i$. The fractional residue of each ratio resulting from the flooring operation is carried over across successive subcycles. Whenever the accumulated fractional residue reaches or exceeds one, an additional allocation is made to the corresponding user in that subcycle.
	
	In the first subcycle, all non-integral ratios round up, since:
	
	\[
	\floor*{\frac{ds^*}{ds_i}} \cdot ds_i \leq ds^* < \floor*{\frac{ds^*}{ds_i}  + 1} \cdot ds_i
	\]
	
	Let's continue with an example. In a system with two demands, whose dominant share ratio is $1.25$, the shortest subcycle will consist of each demand taking $1$ allocation. In this case, the residual accumulation period is four subcycles: after the initial round-up, three subcycles pass before the next round-up, which adds one more allocation for the lower dominant share.

	In cases where the fractional residues accumulate to \textit{exceed} an integer, such as in the case of a ratio of $1.4$, the residual accumulation periods alternate between two and three subcycles. Thus, after the initial round-up, the next round-up occurs two subcycles later, and the following one three subcycles after that.
	
	Ratios which have a decimal part greater than $0.5$, exceed integers more often than not, and thus will mostly round up, and round down in the fewer cases where they fail to. This is why rounding to the nearest integer is the most frequent outcome.
	
	Systems with more than two users, in which all pairwise dominant share ratios have nonzero fractional parts, and not necessarily as nice as the ones presented in the examples above, lead to a cornucopia of subcycles, with differing and chaotically overlapping rounding periods for each demand. Note that in that variability, it is even possible that $l'$ is never actualised due to phases of rounding down not coinciding for all demands.
	
	Luckily, we will not need to identify all of these for approximating DRF allocation.
	
	\section{Precomputed Dominant Resource Fairness}
	\label{pdrf}
	
	Unlike EDRF, PDRF utilises \textit{absolute} demands, rather than the \textit{normalised fractional demands} throughout the computation process. For this reason, we start by redefining $d_{ij}$ notation as the absolute demand of user $i$ for resource $j$. PDRF computes the aggregate resource demand for one subcycle by scaling each user's demand according to the ratio of the maximum dominant share to that user's dominant share. It then determines the maximum continuous number of subcycle iterations, \(k\), by finding the minimum ratio of resource capacity to aggregate subcycle demand.
	
	\[
	k = \min_{j:\,\sum_i d_{ij}>0}
	\left\{
	\frac{r_j}{\sum_i \left( \frac{ds^*}{ds_i} \cdot d_{ij} \right)}
	\right\}
	=
	\frac{1}{ds^*}
	\min_{j:\,\sum_i d_{ij}>0}
	\left\{
	\frac{r_j}{\sum_i \frac{d_{ij}}{ds_i}}
	\right\}
	\]
	
	\noindent The individual allocations for each user, in turn, are given by:
	
	\[
	u_{ij} = \floor*{k \cdot \frac{ds^*}{ds_i}} \cdot d_{ij}
	\]
	
	\noindent for user $i$. The floor function accounts for the indivisibility of demands and resources. Under a perfect-divisibility assumption, it can simply be omitted.
	
	Note that the $1 / ds^*$ in $k$ and $ds^*$ in the second term cancel out nicely to simplify the terms to the following expressions:
	
	\[
	k' = \min_{j:\,\sum_i d_{ij}>0}
	\left\{
	\frac{r_j}{\sum_i \left( \frac{d_{ij}}{ds_i} \right)}
	\right\}
	\]
	
	\noindent and
	
	\[
	u_{ij} = \floor*{\frac{k'}{ds_i}} \cdot d_{ij}
	\]
	
	In our initial experiments, the formulation retaining \(ds^*\) was observed to provide better numerical accuracy and execution performance than the algebraically simplified formulation. We therefore used the unsimplified version in our experiments. However, subsequent independent tests conducted with ChatGPT (GPT-6; OpenAI) neither reproduced this observation nor provided evidence that would falsify it. Since the original experimental data comparing the two formulations were not retained, this discrepancy could not be investigated further.
	
	\begin{algorithm}
	\begin{algorithmic}[1]
		\State $R = \left\langle r_1, r_2, r_3, \dots , r_m \right\rangle$ \Comment{resource reserves}
		\State $U_i = \left\langle 0, 0, \dots, 0 \right\rangle$ for $i=1,\ldots,n$ \Comment{allocation vectors}
		\State $D_i = \left\langle d_{i1}, d_{i2}, d_{i3}, \dots , d_{im} \right\rangle$ for $i=1,\ldots,n$ \Comment{demands}
		\For{$i=1,\ldots,n$}
		\State $ds_i \gets \max_{j=1,\ldots,m}\frac{d_{ij}}{r_j}$
		\EndFor
		\State $ds^* \gets \max_{i=1,\ldots,n} ds_i$
		\State $d_t \gets \left\langle 0,0,\dots,0 \right\rangle$
		\For{$i=1,\ldots,n$}
			\State $d_t \gets d_t + \frac{ds^*}{ds_i}\cdot D_i$ \Comment{scaled aggregate demand}
		\EndFor
		\State $k \gets \min_{j:\,d_{tj}>0} \left\{ \frac{r_j}{d_{tj}} \right\}$ \Comment{maximum subcycle iterations}
		\For{$i=1,\ldots,n$}
			\State $a_i \gets \floor*{k\cdot\frac{ds^*}{ds_i}}$
			\State $U_i \gets a_i\cdot D_i$ \Comment{allocation vector for user $i$}
		\EndFor
		\State \Return $\left\langle U_1,U_2,\ldots,U_n\right\rangle$
	\end{algorithmic}
	\caption{Precomputed Dominant Resource Fairness (PDRF)}
	\label{pseudocode}
\end{algorithm}

	The subsequent tests did not establish a consistent advantage for either formulation. Nevertheless, this does not invalidate the results obtained using the unsimplified formulation, which we therefore retain and present.
	
	The pseudocode of PDRF is presented in Algorithm \ref{pseudocode}. A sample implementation can be accessed at \href{https://github.com/serdarmetin/Precomputed-Dominant-Resource-Fairness}{this GitHub repository} \citep{metin2025pdrf}.

	\section{Mathematical Model and Bounds for PDRF}
	\label{mathematics}
	
	In this section we present a mathematical model for PDRF and provide bounds on the deviation from the DRF allocation.
	
	\subsection{Mathematical Model of PDRF}
	\label{model}
	
	In order to show the approximation bounds, we need to first present a mathematical model for PDRF. Here, we return to the idea of normalised fractional demands put forward in \cite{parkes2015beyond}, but keep with our own notation, while introducing new variables as needed by the argument.
	
	We assume that every user has a nonzero demand vector; that is, for each user $i$, there is at least one resource $j$ for which $d_{ij} > 0$. Accordingly, $ds_i > 0$ for every user $i$, although individual resource demands may be zero.
	
	Let $r_j$ denote the capacity of resource $j$, and let $T_j$ be its normalised aggregate demand, such that:
	
	\[
	T_j = \sum_{i = 1}^{n} \frac{d_{ij}}{ds_i}
	\]
	
	This is slightly different from the normalised aggregate \textit{fractional} demand used in EDRF, which is defined as:
	
	\[
	A_j = \sum_{i = 1}^{n} \frac{d_{ij}}{r_j \cdot ds_i}.
	\]\footnote{In section \ref{edrf}, keeping with the original paper's symbol convention we annotated the summation term as $d_{ij} = fd_{ij}/ds_i$. In this section we are keeping our own symbol convention, so the terms are reformulated accordingly.
	}
	
	In the EDRF model, fractional demands are used throughout derivation. $A_j$ values are reciprocated, to scale the fractional aggregates to $1$, and the minimum $a = (1 / max_j A_j)$\footnote{In the original notation this is represented with $x$.} is used to find the maximum \textit{continuous} allocation. This is useful in the mathematical analysis of DRF, but it does not lead to an allocation scheme where only integer numbers of allocation makes sense. Though the mathematical models can be reduced to each other, and in this respect may said to be identical, the novelty presented here is to operate with \textit{discrete} values, by rounding down the continuous values in the necessary steps. It turns out that it is not a straightforward to apply constructs established in the continuous conditions to discrete conditions.
	
	We continue with defining $\lambda$, $p_i$, and $x_i$ such that:
	
	\[
	\lambda = \min_{j:T_j > 0} \frac{r_j}{T_j}, \quad
	x_i = \frac{\lambda}{ds_i}, \quad
	p_i = \floor{x_i}.
	\]
	
	\noindent omitting the resources which no user submitted a demand for, if any. In this notation, $\lambda$ is equivalent with the reciprocal of the $a$ value used in the EDRF context (i.e. $\lambda = 1 / a$), and $p_i$ corresponds to the number of task allocations for user $i$ by PDRF. In terms employed in Section \ref{pdrf}:
	
	\[
	p_i = \floor*{x_i} = \floor*{\frac{\lambda}{ds_i}} = \floor*{k \cdot \frac{ds^*}{ds_i}}, \quad
	\lambda = k\cdot ds^*
	\]
	
	\subsection{Feasibility of PDRF}
	
	PDRF is feasible, i.e. the total allocations are guaranteed to not exceed the resource reserves. By the definition of the floor function:
	
	\[
	p_i = \floor{x_i} \leq x_i.
	\]
	
	\noindent therefore
	
	\[
	C_j(p) := \sum_{i=1}^{n} p_i\cdot d_{ij}
	\leq \sum_{i=1}^{n} x_i\cdot d_{ij}
	= \lambda \cdot T_j \leq r_j.
	\]
	
	\noindent which concludes that the algorithm is feasible by the definitions of $\lambda$ and the floor function.
	
	\subsection{Bounds for Difference from DRF}
	\label{bounds}
	
	Next, we provide bounds for difference of PDRF allocation from DRF allocation. We start by stating a lemma.
	
	\subsubsection{Lemma}
	\label{lemma}
	
	Let $G = \left\langle g_1, g_2, g_3, \dots, g_n \right\rangle$ denote the DRF allocation, similar to $p_i$ (i.e. allocation of user $i$ for resource $j$ is given by $g_i \cdot d_{ij}$), and let $q_i = g_i \cdot ds_i$ and $h = \min_i q_i$. At each step of allocation the following inequality holds:
	
	\[
	h \leq q_i \leq h + ds_i
	\]
	
	\noindent for every user $i$. In other words, no user can have more than one of her dominant share away from $h$, according to the right side of the inequality. This is ensured by the fact that DRF makes allocations to the user with the minimum allocated total dominant share (i.e. $h$). Therefore, immediately before the allocation, the user has $h$, and after the allocation, $h+ds_i$. At this point if there are other users with $h$, they take turn, according to the predefined tie-breaking policy. When they are also allocated, $h$ updates to the next minimum, which, by definition, can only be higher than the previous value. Since $h$ never decreases, and no user can get allocations unless their total allocated dominant share equals to $h$, the right hand side of the inequality holds. The left side is the case according to the definition of $h$.
	
	We can show that $h \leq \lambda$:
	
	\[
	h\cdot T_j \leq C_j(g) \leq r_j \implies h \leq \frac{r_j}{T_j}
	\]
	
	\noindent and
	
	\[
	\lambda = \min_{j:T_j > 0} \frac{r_j}{T_j} \implies h \leq \lambda
	\]
	
	Now let user $u$ cause the allocation to terminate because there is not enough capacity in resource $j_0$ to allocate. Let $h = q_u = g_u \cdot ds_u$ denote the minimum allocated dominant share immediately before the failed allocation. We have:
	
	\[
	g_i\cdot ds_i \leq h + ds_i
	\]
	
	\noindent dividing each side with $ds_i > 0$, we obtain
	
	\[
	g_i \leq \frac{h}{ds_i} + 1
	\]
	
	\noindent Since demands are non-negative, multiplying each side with $d_{ij_0}$ gives us
	
	\[
	g_i \cdot d_{ij_0}
	\leq \frac{h \cdot d_{ij_0}}{ds_i} + d_{ij_0}
	\]
	
	We now compute the sum of the expression above over all users except $u$ and handle $u$ separately, since we will later use specific equality $h = q_u$ and the corollary $h / ds_u = g_u$ at the time of termination, rather than the general inequality $h \leq q_i \leq h + ds_i$.
	
	\[
	\sum_{i \neq u} g_i \cdot d_{ij_0}
	\leq h \sum_{i \neq u} \frac{d_{ij_0}}{ds_i}
	+ \sum_{i \neq u} d_{ij_0}
	\]
	
	Turning to $u$:
	
	\[
	g_u = \frac{h}{ds_u}
	\implies
	\left( g_u + 1 \right) \cdot d_{uj_0}
	= \left(\frac{h}{ds_u} + 1\right) \cdot d_{uj_0}
	\]
	
	The capacity of resource $j_0$ needed for adding one more task to user $u$ is given by combining the two expressions:
	
	\[
	\begin{aligned}
		&C_{j_0} (g) + d_{uj_0}
		= \sum_{i \neq u} g_i \cdot d_{ij_0}
		+ \left(g_u + 1\right) \cdot d_{uj_0}\\
		&\leq h \sum_{i \neq u} \frac{d_{ij_0}}{ds_i}
		+ \sum_{i \neq u} d_{ij_0}
		+ \left(g_u + 1\right) \cdot d_{uj_0}\\
		&= h \sum_{i \neq u} \frac{d_{ij_0}}{ds_i}
		+ \sum_{i \neq u} d_{ij_0}
		+ \left( \frac{h}{ds_u} + 1 \right) \cdot d_{uj_0}
	\end{aligned}
	\]
	
	\noindent by integrating the $u$ terms into the summations over all users on the right hand side of the equation we obtain:
	
	\[
	C_{j_0} (g) + d_{uj_0}
	\leq h \cdot \sum_{i = 1}^{n} \frac{d_{ij_0}}{ds_i}
	+ \sum_{i = 1}^{n} d_{ij_0}
	= h \cdot T_{j_0} + \sum_{i = 1}^{n} d_{ij_0}
	\]
	
	\noindent since by definition $ds_i \leq ds^*, \forall i$, it is obvious that $1 \leq ds^* /ds_i \forall i$
	
	\[
	d_{ij_0} \leq \frac{ds^*}{ds_i} \cdot d_{ij_0}
	\implies
	\sum_{i = 1}^{n} d_{ij_0}
	\leq ds^* \cdot \sum_{i=1}^{n}\frac{d_{ij_0}}{ds_i}
	= ds^* \cdot T_{j_0}
	\]
	
	\noindent replacing the last term of the penultimate inequality with the bound we obtained in the last derivation and regrouping the terms on the right side yields:
	
	\[
	C_{j_0}(g) + d_{uj_0}
	\leq \left( h + ds^*\right) \cdot T_{j_0}
	\]
	
	Since by construction the additional allocation is not feasible, we may add the following inequality to the left hand side to obtain:
	
	\[
	\lambda \cdot T_{j_0} \leq r_{j_0}
	< C_{j_0}(g) + d_{uj_0}
	\leq \left( h + ds^*\right) \cdot T_{j_0}
	\]
	
	\noindent removing the middle terms and simplifying the normalised aggregate demand $T_{j_0}$ gives:
	
	\[
	\lambda < h + ds^* \implies \lambda - ds^* < h
	\]
	
	\noindent combining with the inequality $h \leq \lambda$ results in the final form:
	
	\[
	\lambda - ds^* < h \leq \lambda
	\]
	
	\subsubsection{Bound Theorem}
	
	What we intend to show is that PDRF allocation ($p_i$) differs from DRF ($g_i$), by at most one allocation fewer, or $ds^*/ds_i$ allocation more for any user $i$. That is to say:
	
	\[
	-1 \leq p_i - g_i \leq \frac{ds^*}{ds_i}
	\]
	
	We start by showing the lower bound. As shown in Lemma \ref{lemma}, any user can get at most one dominant share more from the allocated minimum total dominant share given to any user (i.e. $q_i \leq h + ds_i$), and also $h \leq \lambda$. Combining the two we get:
	
	\[
	q_i \leq h + ds_i \leq \lambda + ds_i
	\]
	
	\noindent since $q_i$ is defined as $q_i = g_i \cdot ds_i$, removing the middle term and dividing each side by $ds_i$ gives:
	
	\[
	g_i \leq \frac{\lambda}{ds_i} + 1
	\]
	
	\noindent by the fact that $g_i$ is an integer (number of allocated tasks by DRF) it follows that
	
	\[
	g_i \leq \floor*{\frac{\lambda}{ds_i}} + 1 = p_i + 1
	\implies -1 \leq p_i - g_i
	\]
	
	For showing the upper bound we use the lower bound of the final inequality of Lemma \ref{lemma}, which states that $h > \lambda - ds^*$ and we also have $g_i \cdot ds_i \geq h$, as we have recently used above. Combining the two yields $g_i \cdot ds_i > \lambda - ds^*$.
	
	In addition to that, by definition $p_i$:
	
	\[
	p_i = \floor*{\lambda / ds_i}
	\implies p_i \cdot ds_i \leq \lambda
	\]
	
	\noindent subtracting the two inequalities side by side:
	
	\[
	p_i\cdot ds_i - g_i \cdot ds_i
	< \lambda - \left(\lambda - ds^*\right) = ds^*
	\]
	
	\noindent Finally, factoring out $ds_i$ terms on the left and dividing both sides by it we obtain the upper bound:
	
	\[
	\left(p_i - g_i\right)\cdot ds_i < ds^*
	\implies p_i - g_i < \frac{ds^*}{ds_i}
	\]
	
	The combined form of upper and lower bounds shows the theorem. Since both $g_i$ and $p_i$ are integer numbers, a slightly stricter upper bound and the conclusion is:
	
	\[
	\boxed{-1 \leq p_i-g_i \leq \ceil*{\frac{ds^*}{ds_i}}-1
	}.
	\]
	
	\subsubsection{No Uniform Constant Upper Bound}
	
	The theorem above proves the lower bound of discrepancy between the PDRF allocation and DRF allocation is bounded below by $-1$, yet leaves the upper bound for each user demand $i$ dependent on the ratio of the highest dominant share over the dominant share of the given user demand (i.e. $ds^* / ds_i$). Since the ratio may be arbitrarily large, we look for a better bound; if possible independent of the user input.
	
	Unfortunately, it turns out that no such bound can exist. The proof is simply by a counterexample.
	
	Let a system with resources $R = \left\langle r_1, r_2 \right\rangle$, with capacities $r_1 = Z$ and $r_2 = 2Z - 1$, serve users A and B, with demand vectors $d_a = \left\langle 1, 1 \right\rangle$ and $d_b = \left\langle 1, Z \right\rangle$, respectively, for some $Z = 2z \geq 4, Z,z \in \mathbb{N}$. The dominant shares of the users are:
	
	\[
	ds_a = \frac{1}{Z}, \qquad ds_b = \frac{Z}{2Z - 1}
	\]
	
	\noindent respectively.
	
	We first derive the DRF allocation. Consistent with the notation used so far, let \(g_i\) denote the \(i^{th}\) component of the vector \(G\), representing the total task allocation, and let \(q_i\) denote the \(i^{th}\) component of the vector \(Q\), representing the total dominant resource allocated to user \(i\). In particular, \(q_i\) is given by \(q_i=g_i\cdot ds_i\). According to its main working principle, DRF allocates $1$ task to the user with lowest $q_i$ (defined as $h$ in Lemma \ref{lemma}), and when there are more than one user with $q_i = h$, it applies a tie-breaking policy. For this example it does not matter but let us take the convention of choosing the user with the greater dominant share employed in \cite{ghodsi2011dominant}. At the beginning of the allocation $Q = \left\langle 0, 0 \right\rangle$, so the algorithm chooses User B and allocates 1 task. After that $Q$ updates to $Q = \left\langle 0, Z / (2Z - 1) \right\rangle$.
	
	Observing that $Z$ is an even number, and at least $4$
	
	\[
	q_b = \frac{Z}{2Z - 1} = \frac{2z}{4z - 1} > \frac{1}{2}
	\]
	
	This results in $q_a = h$, and the algorithm starts repeatedly selecting User A, until $q_b \leq q_a$ is achieved. Since $ds_a = 1/Z = 1/(2z)$, allocating $j$ tasks to User A increases $q_a$ to $j / 2z$. When $j = z$:
	
	\[
	q_a = z\cdot\frac{1}{2z} = \frac{1}{2}
	\]
	
	At which point still $q_b > q_a$, so User A is selected again. In the next allocation,
	
	\[
	q_a = (z + 1) \cdot \frac{1}{2z} = \frac{z + 1}{2z}
	\]
	
	\noindent When we compare the two we find that
	
	\[
	\frac{z+1}{2z} > \frac{2z}{4z-1}
	\implies (z+1)\cdot(4z-1) > 4z^2
	\implies (3z - 1 > 0)
	\]
	
	\noindent the resulting inequality holds for $z \geq 2$, which is satisfied by our choosing of $Z \geq 4, Z = 2z$.
	
	So for the next allocation User B is selected. But User B cannot be allocated because a second allocation to her will need $2Z$ of $r_2$ capacity. At this point the algorithm terminates and the final allocation is:
	
	\[
	\boxed{g_a = z + 1 = \frac{Z}{2} + 1, \qquad g_b = 1}
	\]
	
	Now we can derive PDRF allocation. PDRF starts with finding $ds^* = ds_b$, and obtains the ratios of each dominant share to $ds^*$. Therefore:
	
	\[
	\frac{ds^*}{ds_a}
	= \frac{\frac{Z}{2Z - 1}}{\frac{1}{Z}}
	= \frac{Z^2}{2Z - 1}, \qquad
	\frac{ds^*}{ds_b} = 1
	\]
	
	The capacity exhaustion for each resource in one subcycle is calculated, and the resources are divided by the resulting values, to take the minimum to serve as the maximum number of possible iterations. This can be reformulated, as shown in Section \ref{pdrf}, as
	
	\[
	k = \min_{j=1,\dots,m}
	\left\{
	\frac{r_j}{\sum_i \left( \frac{ds^*}{ds_i} \cdot d_{ij} \right)}
	\right\}
	= \frac{1}{ds^*}
	\min_{j=1,\dots,m}
	\left\{
	\frac{r_j}{\sum_i \frac{d_{ij}}{ds_i}}
	\right\}
	= \frac{1}{ds^*}
	\min_{j:T_j>0}
	\left\{
	\frac{r_j}{T_j}
	\right\}
	\]
	
	So what we get when we divide the resource capacities by the scaled sums can be notated as $ds^* \cdot T_j$, and they are calculated as:
	
	\[
	ds^* \cdot T_1
	= \frac{Z^2}{2Z-1}+1
	= \frac{Z^2+2Z-1}{2Z-1},
	\]
	
	\[
	ds^* \cdot T_2
	= \frac{Z^2}{2Z-1}+Z
	= \frac{3Z^2-Z}{2Z-1}.
	\]
	
	The maximum available number of subcycles are given by:
	
	\[
	\begin{aligned}
		k
		&=\min\left\{\frac{Z}{ds^* \cdot T_1},
		\frac{2Z-1}{ds^* \cdot T_2}\right\}\\[3pt]
		&=\min\left\{
		\frac{Z(2Z-1)}{Z^2+2Z-1},
		\frac{(2Z-1)^2}{3Z^2-Z}
		\right\}.
	\end{aligned}
	\]
	
	For $Z \geq 4$ the second expression is smaller. So
	
	\[
	\boxed{k=\frac{(2Z-1)^2}{3Z^2-Z}.}
	\]
	
	The allocations are given as
	
	\[
	\boxed{
		p_i = \floor*{k \cdot \frac{ds^*}{ds_i}} \implies
		p_a = \floor*{\frac{Z(2Z-1)}{3Z-1}}, \quad
		p_b = \floor*{k} = 1
	}
	\]
	
	When we calculate the difference between the PDRF allocation and DRF allocation of User A (i.e. $p_a - g_a$) we find:
	
	\[
	p_a - g_a
	= \floor*{\frac{Z(2Z-1)}{3Z-1}}
	- \left( \frac{Z}{2} + 1 \right)
	\]
	
	We use the inequality $\floor*{x} > x - 1$ to get rid of the floor function and end up with the following inequality:
	
	\[
	p_a - g_a
	> \frac{Z(2Z-1)}{3Z-1} - 1
	- \left( \frac{Z}{2} + 1 \right)
	= \frac{Z(Z - 1)}{2(3Z -1)} - 2
	\]
	
	Therefore:
	
	\[
	\boxed{
		p_a - g_a > \frac{Z(Z - 1)}{2(3Z -1)} - 2
	}
	\]
	
	When we inspect the limit of this expression as $Z$ goes to infinity we find that:
	
	\[
	\frac{Z(Z - 1)}{2(3Z -1)} - 2
	\approx \frac{Z}{6}, \quad \text{ as } Z \to \infty
	\]
	
	\noindent Thus:
	
	\[
	\boxed{
		\lim_{Z \to \infty}
		\left[\frac{Z(Z - 1)}{2(3Z -1)} - 2\right] = \infty
	}
	\]
	
	This concludes that no finite, input-independent upper bound exists for the divergence of PDRF from DRF.

	\section{Experimental Findings}
	\label{experiments}
	
	In this section we present the results of the tests we have carried out by samples drawn from the discrete uniform distribution.
	
	$1000$ users are assigned random demands, and resource reserves are likewise assigned randomly. We tested for various numbers of resource types, demand intervals and resource reserve intervals. Here, we are presenting a sample which we consider representative. We run each test $1000$ times, and collected the averages, standard deviations, and maximums, presented in Tables \ref{table:table_1} - \ref{table:table_4}. The scripts that we used to run the tests can be accessed at \href{https://github.com/serdarmetin/Precomputed-Dominant-Resource-Fairness}{the same repository, here} \citep{metin2025pdrf}.
	
	The results show that the deviations of the PDRF allocation from the DRF allocation are limited to a small number of tasks per user, predominantly by $1$, and very rarely exceeding $2$. However, the error is two tailed, meaning, the deviations may either be underallocations or overallocations, the latter of which is a rather serious drawback, since, unlike underallocations, they cannot be compensated with additional allocations. Nevertheless, the rate at which these occur are negligibly low. 
	
	In the tests carried out with resource intervals drawn from $[50,000-100,000]$ interval and demands are drawn from $[1-10]$ interval, on avarage, $47.7\%$ ($\mu = 477.623$, $n = 1000$) of the users are underallocated $1$ task, with a standard deviation of $\sigma = 86.373$. The figures for demand intervals $[1-20]$ and $[10-20]$ are similarly, $\mu = 474.932$, $\sigma = 38.381$ and $\mu = 486.521$, $\sigma = 50.526$, respectively. The overallocations are for demand intervals $[1 - 10]$ and $[1 - 20]$ are $\mu = 0.694$, $\sigma = 2.137$ and $\mu = 0.192$, $\sigma = 0.526$, respectively.

\begin{figure}
		\centering
		\includegraphics[width=1\linewidth]{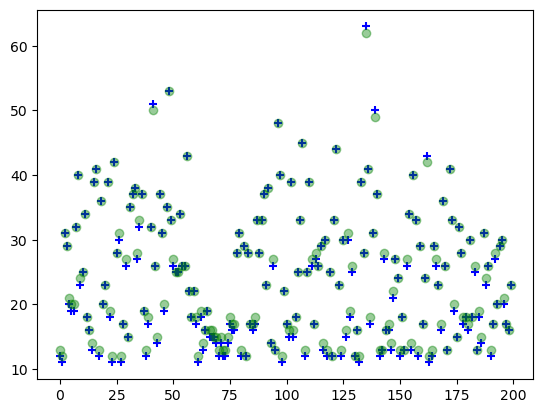}
		\caption{Scatterplot for number of allocations made by DRF (circles) and PDRF (crosses), with $200$ users and $10$ resources.}
		\label{scatterplot}
	\end{figure}

	Figure \ref{scatterplot} shows a sample case for the number of tasks allocated by DRF and PDRF in an overlapping scatterplot. The tests is run with $200$ users for the clarity of the view, from the demand interval $1-100$, with $10$ resources, each resource drawn from $100000-1000001$ interval.

\begin{table}[h!]
	\resizebox{\columnwidth}{!}{
		\begin{tabular}{|c|c|c|c|c|c|c|}
			\hline
			Demand & 1 task & 2 tasks & More & Max. & Avg. & St. Dev. \\
			\hline
			1 – 10 & 477.623 & 0 & 0 & 1 & 477.623 & 86.373 \\
			\hline
			1 – 20 & 474.932 & 0 & 0 & 1 & 474.932 & 38.381 \\
			\hline
			10 – 20 & 486.521 & 0 & 0 & 1 & 486.521 & 50.526 \\
			\hline
		\end{tabular}
	}
	\caption{Average Case Scenarios - Underallocations: Tests Done with 1000 users, 10 resources, Resource Reserves Drawn From [50,000 – 100,000] Interval}
	\label{table:table_1}
\end{table}

\begin{table}[h!]
	\resizebox{\columnwidth}{!}{
		\begin{tabular}{|c|c|c|c|c|c|c|}
			\hline
			Demand & 1 task & 2 tasks & More & Max. & Avg. & St. Dev. \\
			\hline
			1 – 10 & 0.693 & 0.001 & 0 & 2 & 0.694 & 2.137 \\
			\hline
			1 – 20 & 0.191 & 0.001 & 0 & 2 & 0.192 & 0.526 \\
			\hline
			10 – 20 & 0 & 0 & 0 & 0 & 0 & 0 \\
			\hline
		\end{tabular}
	}
	\caption{Average Case Scenarios - Overallocations: Tests Done with 1000 users, 10 resources, Resource Reserves Drawn From [50,000 – 100,000] Interval}
	\label{table:table_2}
\end{table}

	An observation at this point is that the overallocations are biased towards the demands with lower dominant shares, especially the demands containing $1$, with demands of $1$ for all resource types being the most falsely favoured. As seen in Table \ref{table:table_2}, when the lower end of the interval is moved to $10$, overallocations are totally avoided.
	
	\begin{figure}[H]
		\centering
		\includegraphics[width=1\linewidth]{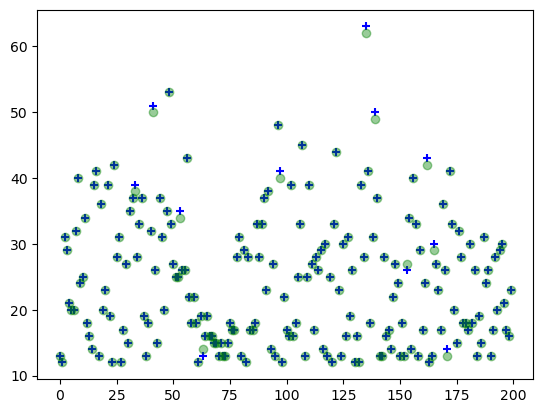}
		\caption{Scatterplot for number of allocations made by DRF (circles) and PDRF with heuristic method (crosses), with $200$ users and $10$ resources.}
		\label{heuristic_scatterplot}
	\end{figure}
	
	In more general terms, for a given demand, as the ratio of maximum dominant share to its dominant share increases, the number of overallocations increase proportionally. This is caused by large demand and resource reserve intervals which lead to increased variances in the distribution of dominant shares, and mitigated with the increasing number of resource types, since it leads to lower variance among dominant shares. The worst configuration we could cook were with $2$ resource types, as seen in Tables \ref{table:table_3} and \ref{table:table_4}.
	
	The reason for the algorithm's implicit favouring of the lower dominant shares is due to the the rather abstract notion of \textit{partial cycles}, which are reflected in the decimal points of the number of cycle iterations. If the exact ordering of the cycle were available, those numbers would indicate the fraction of the cycle that would execute from the beginning, e.g. a decimal point of 0.5 would indicate that the first half of the allocations would be done and the cycle would terminate. In the absence of the exact ordering of allocations, the users exploit the partial cycles inversely proportional to the ratio of the $ds^*$ to their $ds$.

	\begin{table}[h!]
		\resizebox{\columnwidth}{!}{
			\begin{tabular}{|c|c|c|c|c|c|c|}
				\hline
				Demand & 1 task & 2 tasks & More & Max. & Avg. & St. Dev. \\
				\hline
				1-100 & 428.653 & 0 & 0 & 1 & 428.653 & 21.919 \\
				\hline
				1-1000 & 430.86 & 0 & 0 & 1 & 430.86 & 37.002 \\
				\hline
			\end{tabular}
		}
		\caption{Worst Case Scenarios - Underallocations: Tests Done with 1000 users, 2 resources, Resource Reserves Drawn From [500,000 – 1,000,000] Interval}
		\label{table:table_3}
	\end{table}

	\begin{table}[h!]
		\resizebox{\columnwidth}{!}{
			\begin{tabular}{|c|c|c|c|c|c|c|}
				\hline
				Demand & 1 task & 2 tasks & More & Max. & Avg. & St. Dev. \\
				\hline
				1-100 & 40.267 & 10.12 & 9.239 & 35 & 59.626 & 10.790 \\
				\hline
				1-1000 & 33.061 & 14.519 & 8.256 & 610 & 49.819 & 27.055 \\
				\hline
			\end{tabular}
		}
		\caption{Worst Case Scenarios - Overallocations: Tests Done with 1000 users, 2 resources, Resource Reserves Drawn From [500,000 – 1,000,000] Interval}
		\label{table:table_4}
	\end{table}
	
	\section{An Heuristic Method for Allocating Remaining Resources}
	\label{heuristic}
	
	Based on the observations we made from the results, a simple heuristic approach to improve accuracy is to allocate $1$ task to each user, in the ascending order of allocated dominant shares, to as much users as possible, since the number of underallocations is invariably by $1$.
	
	With this extra step, the average underallocation rate across the three demand configurations reported in Tables \ref{table:table_5} and \ref{table:table_6} falls to $\mu = 1.2\%$, with a standard deviation of $\sigma = .559$, while the average overallocation rate rises to $\mu = 1.3\%$, with a standard deviation of $\sigma = .543$. In the worst-case configurations reported in Tables \ref{table:table_7} and \ref{table:table_8}, the corresponding average rates are $\mu = 1.6\%$ ($\sigma = .257$) and $\mu = 5.8\%$ ($\sigma = .638$), respectively. The results with the addition of the heuristic method are presented in more detail in Tables \ref{table:table_5}–\ref{table:table_8}.	
	
	Nevertheless it should be noted that this heuristic method is specific to the discrete uniform distribution, and there is no guarantee it will work under different distributions.

	\begin{table}[h!]
		\resizebox{\columnwidth}{!}{
			\begin{tabular}{|c|c|c|c|c|c|c|}
				\hline
				Demand & 1 task & 2 tasks & More & Max. & Avg. & St. Dev. \\
				\hline
				1 – 10 & 17.600 & 0 & 0 & 1 & 17.600 & 15.114 \\
				\hline
				1 – 20 & 6.436 & 0 & 0 & 1 & 6.436 & 5.937 \\
				\hline
				10 – 20 & 11.459 & 0 & 0 & 1 & 11.459 & 8.710 \\
				\hline
			\end{tabular}
		}
		\caption{Average Case Scenarios - Underallocations: Tests Done with 1000 users, 10 resources, Resource Reserves Drawn From [50,000 – 100,000] Interval}
		\label{table:table_5}
	\end{table}
	
	\begin{table}[h!]
		\resizebox{\columnwidth}{!}{
			\begin{tabular}{|c|c|c|c|c|c|c|}
				\hline
				Demand & 1 task & 2 tasks & More & Max. & Avg. & St. Dev. \\
				\hline
				1 – 10 & 19.116 & 0 & 0 & 1 & 19.116 & 15.266 \\
				\hline
				1 – 20 & 8.521 & 0 & 0 & 1 & 8.521 & 6.160 \\
				\hline
				10 – 20 & 11.716 & 0 & 0 & 1 & 11.716 & 8.764 \\
				\hline
			\end{tabular}
		}
		\caption{Average Case Scenarios - Overallocations: Tests Done with 1000 users, 10 resources, Resource Reserves Drawn From [50,000 – 100,000] Interval}
		\label{table:table_6}
	\end{table}

	\begin{table}[h!]
		\resizebox{\columnwidth}{!}{
			\begin{tabular}{|c|c|c|c|c|c|c|}
				\hline
				Demand & 1 task & 2 tasks & More & Max. & Avg. & St. Dev. \\
				\hline
				1-100 & 18.412 & 0 & 0 & 1 & 18.412 & 5.694 \\
				\hline
				1-1000 & 14.773 & 0 & 0 & 1 & 14.773 & 9.928 \\
				\hline
			\end{tabular}
		}
		\caption{Worst Case Scenarios - Underallocations: Tests Done with 1000 users, 2 resources, Resource Reserves Drawn From [500,000 – 1,000,000] Interval}
		\label{table:table_7}
	\end{table}

	\begin{table}[h!]
		\resizebox{\columnwidth}{!}{
			\begin{tabular}{|c|c|c|c|c|c|c|}
				\hline
				Demand & 1 task & 2 tasks & More & Max. & Avg. & St. Dev. \\
				\hline
				1-100 & 43.608 & 9.71 & 9.53 & 34 & 62.848 & 11.139 \\
				\hline
				1-1000 & 36.698 & 8.413 & 8.714 & 259 & 53.825 & 28.084 \\
				\hline
			\end{tabular}
		}
		\caption{Worst Case Scenarios - Overallocations: Tests Done with 1000 users, 2 resources, Resource Reserves Drawn From [500,000 – 1,000,000] Interval}
		\label{table:table_8}
	\end{table}
	
	As an additional consistency check, the experiments were independently rerun using ChatGPT (GPT-6; OpenAI). The resulting statistics were similar to those reported here, providing an independent indication of the consistency of the experimental procedure and its implementation. The results of the GPT experiments can be accessed at the same repository \cite{metin2025pdrf}.

	\section{Iteration Complexity of DRF}
	\label{drf_complexity}
	
	We now show that the number of allocation operations in the original DRF algorithm depends on the numerical values of resource capacities and task demands. We establish bounds on this number to characterise DRF iteration complexity, to compare overall DRF complexity with PDRF complexity, in the next section.
	
	\subsection{Allocation Ordering Property}
	
	Before a user with maximum dominant share $ds^*$ receives an additional allocation, every other user must have received at least as many allocations as that user. This is another property emerging from the characterisation of a subcycle.
	
	To see that, let $u_j$ be a user with dominant share $ds_j=ds^*$, and let $g_j'$ denote the number of allocations they have received at the current stage of the algorithm. Here, the prime indicates a current allocation count rather than the final allocation count. Suppose that another user $i$ has received fewer allocations, so that $g_i'<g_j'$, yet user $u_j$ is selected for the next allocation. Their allocated dominant shares then satisfy
	\[
	g_i' ds_i < g_j' ds_i \leq g_j' ds^*.
	\]
	
	\noindent This contradicts the DRF selection rule, which selects the user with the smallest allocated dominant share. Therefore, every other user must have received at least as many allocations as user $u_j$ before they can receive another allocation.
	
	\subsection{Bounds on the Total Number of Allocations}
	
	Let $K=\sum_i g_i$ denote the total number of successful allocation operations before DRF terminates. For each resource $j$, let $M_j=\max_i d_{ij}$, and let $r_{\max}=\max_j r_j$. Then:
	\[
	\boxed{
		\left\lfloor\min_{j:M_j>0}\frac{r_j}{M_j}\right\rfloor
		\leq K \leq nr_{\max}.
	}
	\]
	
	We can show the upper and lower bounds of the inequality above separately.
	
	For the upper bound, consider any user $i$ and their dominant resource $j$, corresponding to $ds_i$. Since the allocation must remain feasible,
	\[
	g_i d_{ij}\leq C_j(g)\leq r_j.
	\]
	Because demands and capacities are integers and $d_{ij}\geq 1$, it follows that
	\[
	g_i\leq \left\lfloor\frac{r_j}{d_{ij}}\right\rfloor
	\leq r_{\max}.
	\]
	Summing over all users gives
	\[
	K=\sum_i g_i\leq nr_{\max}.
	\]
	
	For the lower bound, let $j_0$ be a resource that prevents the next allocation to the selected user $u$. By the termination condition,
	\[
	C_{j_0}(g)+d_{u j_0}>r_{j_0}.
	\]
	By definition, $d_{u j_0}\leq M_{j_0}$. Furthermore, each successful allocation consumes at most $M_{j_0}$ units of resource $j_0$, so
	\[
	C_{j_0}(g)\leq KM_{j_0}.
	\]
	Combining these inequalities gives
	\[
	KM_{j_0}>r_{j_0}-d_{u j_0}
	\geq r_{j_0}-M_{j_0},
	\]
	and consequently
	\[
	K>\frac{r_{j_0}}{M_{j_0}}-1.
	\]
	Since $K$ is an integer, this implies
	\[
	K\geq\left\lfloor\frac{r_{j_0}}{M_{j_0}}\right\rfloor
	\geq
	\left\lfloor\min_{j:M_j>0}\frac{r_j}{M_j}\right\rfloor.
	\]
	This completes the proof.
	
	\subsection{Implications for Computational Complexity}
	
	The upper bound establishes that the number of allocation operations is polynomially bounded in the numerical values of the input. The lower bound shows that this number can grow with the numerical ratio between resource capacities and task demands. For example, consider a single-resource instance with capacity $Q$, where all $n$ users have identical unit demands. DRF performs exactly $Q$ successful allocation operations before termination. Thus, even for a fixed number of users and resource types, the number of operations can grow linearly with a numerical input value. Since the binary encoding length of $Q$ is $\Theta(\log Q)$, the number of allocation operations can grow exponentially with the encoding length of the capacity. Since the upper bound is polynomial in the numerical input values, the iteration complexity of DRF is pseudo-polynomial rather than polynomial in the input encoding length.

	\section{Complexity Analysis of DRF and PDRF}
	\label{performance}
	
	In the original DRF algorithm, Ghodsi et al. state that each scheduling decision takes $\mathcal{O}(\log n)$ time when a binary heap is used. However, this accounts only for selection of the user with the smallest allocated dominant share, but not for checking whether the next task can be allocated, which requires examining the $m$ resource constraints. A single allocation operation, therefore, takes $\mathcal{O}(m+\log n)$ time.	
	
	As shown in Section \ref{drf_complexity}, the number of allocation operations, on the other hand, depends on the numerical values of resource capacities and task demands, rather than solely on the number of users and resource types. Consequently, the total running time cannot be characterised by the cost of a single allocation operation alone. The number of iterations depends on numerical magnitudes whose encoding lengths may be comparatively small, so the algorithm has pseudo-polynomial running time.
	
	The main performance contribution of PDRF is that, in such cases, it can reduce the execution time substantially. Unlike the original DRF algorithm, the running time of PDRF does not depend on the numerical magnitudes of resource capacities and task demands. This distinction is particularly relevant when the ratio of resource capacities to task demands is large, a common feature of cloud-computing scenarios in which clusters offer substantial resources to users whose tasks demand relatively small amounts. Such instances may require DRF to perform many allocation iterations. Partitioning the cluster and the demand set may mitigate this problem by improving the scale of the allocation problem. However, partitioning can only mitigate the problem and it introduces further optimisation problems, some of which are discussed in \cite{ghodsi2011dominant}.
	
	The PDRF core requires $\mathcal{O}(m \cdot n)$ time: computing the dominant shares, scaling and summing the demand vectors, determining the maximum number of subcycle iterations, and constructing the allocations each require at most a constant number of passes over the users and resources. The heuristic step first sorts the $n$ users by their allocated dominant shares, requiring $\mathcal{O}(n\cdot\log n)$ time, and then checks at most one additional task for each user against the $m$ resource constraints, requiring $\mathcal{O}(m\cdot n)$ time. Thus, the overall time complexity of core PDRF is $\mathcal{O}(m\cdot n)$, and with the heuristic step $\mathcal{O}(m\cdot n+n\cdot \log n)$.
	
	Since scheduling is a real-time process, even small reductions in execution time may be important. Cross-comparative results constructed with real data may provide a more accurate measurement of the practical speed-up, but at present we cannot provide such results because we do not have access to these data.
		
	\section{Discussion}
	\label{discussion}

	\subsection{DRF is not Strategy-Proof}
	\label{strategy}
	
	A notorious problem with DRF is that it can be hijacked simply by submitting inflated demands. Consider users A and B, running the same task, with the same resource requirements, $d$. Suppose the total available resources are $4d$. If both users submit their actual demands, DRF allocates two tasks to each, giving each user an allocation of $2d$. Now suppose User B submits $3d$ instead of $d$. The allocation procedure can then allocate one task to A and one task to B, exhausting the available resources. User A receives $d$, whereas User B receives $3d$. Since B's actual task still requires only $d$, B can use this allocation to execute three tasks rather than two. Thus, B benefits by submitting an inflated demand. More generally, users may have an incentive to submit integer multiples of their actual demand vectors, and the reported demands can be arbitrarily inflated.
	
	Ghodsi et al. claim to establish the strategy-proofness of DRF, but the proof they provide relies on a continuous mathematical model. The continuous formulation is an analytical abstraction; DRF itself allocates indivisible tasks, and fractions of tasks do not constitute meaningful task allocations or utility. The counterexample above demonstrates that the strategy-proofness claim does not hold for the original discrete allocation procedure. The issue is not an inconsistency within the proof, but a failure of the proof to establish the claimed property for the mechanism it is intended to analyse. In this sense, the proof is internally consistent, but lacks external validity.
	
	Ghodsi et al. also discuss demand inflation in which a user increases the reported demand for selected resources, rather than scaling the entire demand vector. They also provide an informal argument addressing this strategy in the Competitive Equilibrium from Equal Incomes (CEEI), another multi-resource allocation scheme they use as a benchmark. The counterexample above concerns a different manipulation: reporting an integer multiple of the user's entire demand vector. It demonstrates that the strategy-proofness claim does not hold for the DRF allocation procedure, notwithstanding the proof based on a continuous mathematical abstraction.
	
	This problem creates a conflict between the individual user's incentive and the collective interest in fair allocation. In the example, B benefits by reporting an inflated demand, while A receives less than under truthful reports. In more extreme cases, inflated reports may reduce the number of users whose tasks can be accommodated. Detecting such behaviour may also be difficult, since determining whether a reported demand corresponds to the actual resource requirements of a task may require inspecting the task itself.
	
	The same example also illustrates limitations with respect to envy-freeness and sharing incentive. Under the users' actual task requirements, A receives $d$ while B receives $3d$, so A would prefer B's allocation. The resulting allocation is therefore not envy-free under this utility interpretation. Moreover, an equal partition of the $4d$ resource pool would give each user $2d$, whereas A receives only $d$ under B's inflated report. Hence the example violates the individual sharing-incentive guarantee for A. These observations show that these properties are not guaranteed by DRF in every instance.
	
	PDRF, on the other hand, is immune to this strategy, since it favours smaller dominant shares. A caveat is that demand deflation may present a separate concern. Although deflation is not as straightforward as inflating the entire demand vector by an integer multiple, since tasks are assumed to be indivisible, a user might attempt to report a fractional demand and then execute tasks on the allocated resources while wasting the excess capacity. A simple safeguard is to declare a lower limit for the volume of submittable resource demands, or to define such a limit and internally scale each demand vector up to the smallest integer multiple that meets it. A similar safeguard could be proposed for DRF by introducing an upper limit to demand volumes. However, this would prevent genuine tasks with resource requirements above the upper limit from being run on the server.
	
	The lower limit may also be tailored to minimise the upper bound of divergence established in Section \ref{bounds}, potentially making it arbitrarily small, since this bound is determined by $ds^* / ds_i$. There is, however, a trade-off. Although scaling small demands up, regularises the final allocation, it does so at the cost of the granularity offered to small demands. The resulting problem is an optimisation problem, which needs further scrutiny, and as such, is outside the scope of the present study; we therefore leave it open.
	
	\subsection{Weighting Policies}
	\label{weighting}
		
		A weighting policy, assigns weights to users, and allocates the resources in proportion to these weights. Fairness, in turn, is defined over the allocations scaled by these weights. In \cite{ghodsi2011dominant}, the question of weighting is addressed briefly, pointing to the fact that it suffices for such a generalisation to assign a weight vector to each user, $W_i = \left\langle w_{i1}, w_{i2}, w_{i3}, \dots ,w_{im}\right\rangle$, and compute the dominant shares with $ds_i = \max_j \left\lbrace u_{ij}/w_{ij} \right\rbrace $. In \cite{parkes2015beyond}, weights are further defined to be normalised, such that $\sum_i w_{ir} = 1$, for each resource $r$, and they use another "$\rho$" variable, at the construction of which the weights scale the \textit{normalised demands}, to be used in their calculations. Again, it is notationwise involved and will not be elaborated on in more detail for the sake of simplicity.
		
		In order to implement a weighting policy for PDRF, it suffices to scale the demands with weights for the calculation of dominant shares, as in DRF. On the other hand, as with EDRF, PDRF needs normalised weights, since the selection of $ds^*$, and the proportioning of each $ds$ to it necessitates a standardised measure among them.
		
		At this point, we believe that DRF and its derivatives offer a richer functionality than previously expressed. Even without weights, DRF is an \textit{implicit pricing mechanism}, implicitly assigning \textit{value} to resources on the basis of demand and supply dynamics.
		
		The inclusion of weights articulates this point even stronger. For example, allowing users to \textit{purchase} weight tokens for resources, or open-endedly invest in weights for resources, and assigning their fraction of the total token sum or total investment as the user's weight, DRF can implement an implicit \textit{auctioning} system. Further scrutinising this issue may be a research topic on its own, and is beyond the scope of the present study, thus we will suffice with pointing to it.

	\section{Conclusion}
	\label{conclusion}
	
	The allocation of multiple resources has become an increasingly important problem in the last decade and a half. The seminal study of Ghodsi and his colleagues has created a small literature of its own, since DRF brought many novelties and as much possibilities into this field. The study of Parkes and his colleagues contributed to this literature dearly, by providing a more rigorous theoretical framework to study the properties of DRF in higher resolution.
	
	The present study is one situated in this literature, offering yet another analysis method, and an improvement to the original algorithm to approximate the allocation at high accuracy in fewer steps.
	
	One particular contribution of PDRF is that it is easier to adapt to blockchain context. In fact, it was during our attempt to do so, we came up with the idea, and it lead to the development of a new algorithm for the original context, almost by accident. At that point we turned to a more careful review of the literature, to find EDRF, and the whole endeavour lead to the present study. The blockchain adaptation of PDRF is a forthcoming study.
	
	\section*{Declaration of Generative AI and AI-Assisted Technologies in the Writing Process}
	
	During the preparation of this work, the author utilised ChatGPT (GPT-6; OpenAI), accessed on 27 October 2026, to assist in the formal verification of mathematical models, derivation checks for bounding proofs, and parallel replication of consistency testing routines. Following the use of this technology, the author independently audited, critically reviewed, and verified all mathematical derivations, theorem proofs, and experimental statistics presented in this manuscript. The author takes full responsibility for the technical accuracy, interpretations, and final conclusions of the work.

	\bibliographystyle{elsarticle-num} 
	\bibliography{bibl}
	

	%\begin{thebibliography}{00}
	
	
	%\bibitem{}
	
	%\end{thebibliography}
\end{document}